\documentclass[prd,11pt]{revtex4-1}

\usepackage{amsmath}  

\usepackage{amssymb}
\usepackage{graphicx}   % need for figures
\usepackage{verbatim}   % useful for program listings
\usepackage{color}      % use if color is used in text
\usepackage{subfigure}  % use for side-by-side figures
\usepackage{hyperref}   % use for hypertext links, including those to external documents and URLs
\definecolor{amaranth}{rgb}{0.9, 0.17, 0.31}
\definecolor{purple(munsell)}{rgb}{0.62, 0.0, 0.77}
\definecolor{americanrose}{rgb}{1.0, 0.01, 0.24}
\definecolor{palatinateblue}{rgb}{0.15, 0.23, 0.89}
\definecolor{royalblue(web)}{rgb}{0.25, 0.41, 0.88}
\definecolor{hanpurple}{rgb}{0.32, 0.09, 0.98}
\definecolor{beaublue}{rgb}{0.74, 0.83, 0.9}
\definecolor{carminered}{rgb}{1.0, 0.0, 0.22}
\definecolor{brightpink}{rgb}{1.0, 0.0, 0.5}
\definecolor{vividviolet}{rgb}{0.62, 0.0, 1.0}

\hypersetup{ linktoc=all,
    colorlinks, linkcolor={palatinateblue},
    citecolor={brightpink}, urlcolor={amaranth}}

\newcommand{\changeurlcolor}[1]{\hypersetup{urlcolor=#1}} 

\newcommand{\be}{\begin{equation}}
\newcommand{\ee}{\end{equation}}
\newcommand{\bs}{\begin{split}} 
\newcommand{\bea}{\begin{eqnarray}}
\newcommand{\eea}{\end{eqnarray}}

\begin{document}

\title{Dispersion relations in finite-boost DSR }
%\author{Nosratollah Jafari${}_1$, Michael R.R. Good${}_{1,2}$}
\author{Nosratollah Jafari}\email{nosratollah.jafari@nu.edu.kz }
\author{Michael R.R. Good}
%\affiliation{${}_1$Department of Physics, Nazarbayev University, Astana, Kazakhstan.}
\affiliation{Department of Physics, Nazarbayev University,\\Kabanbay Batyr Ave 53, Nur-Sultan, 010000, Kazakhstan}

%\affiliation{${}^2$Energetic Cosmos Laboratory, Nazarbayev University, Astana, Kazakhstan.}

%

\begin{abstract}
 We  find  finite-boost  transformations  DSR  theories  in  first  order  of  the  Planck  length $l_p$, by solving differential equations for the modified generators. We obtain corresponding dispersion relations for these transformations, which help us classify the DSR theories via four types. The final type of our classification has the same special relativistic dispersion relation but the transformations are not Lorentz. In DSR theories, the velocity of photons is generally different from the ordinary speed $c$ and possess time delay, however in this new DSR light has the same special relativistic speed with no delay. A special case demonstrates that any search for quantum gravity effects in observations which gives a special relativistic dispersion relation is consistent with DSR.

\end{abstract}

\maketitle

\tableofcontents

\newpage

\section{Introduction}

DSR theories have been proposed for quantum gravity (QG) modifications to Einstein's special relativity \cite{ame1,ame2,brun,mag1, mag2}. The transformations of the energy and momentum under finite boosts have been obtained in \cite{Wan}. In this paper, we continue this study by finding the finite-boost DSR transformations to leading order of the Planck length from the solutions of the differential equations, and by looking at the observational consequences of these transformations. We obtain the corresponding dispersion relations, which allow us to classify DSR theories to four types with each type resulting in a different specific kind of DSR to first order of the Planck length.

In fact, the first differential equation for DSR was given by Amelino-Camelia \cite{ame2}. However, despite this paper being one of the pioneering works in DSR, the proposal and differential equations were specific and did not contain all DSR theories to first order. Also, by starting from $\kappa$-Poincar\'e boost generators \cite{Maj}, Amelino-Camelia and colleagues obtained differential equations for a specific DSR theory. They solved the differential equations and found finite transformations for the generators \cite{brun}. These finite transformations are important but they are not the most general. We generalize their method and results.

Maguijio and Smolin obtained a different realization of DSR theories by non-linear action of the Lorentz group on the energy-momentum space \cite{mag1}. Also, they gave a procedure for finding corresponding transformations for any given modified dispersion relation \cite{mag2}. This modified dispersion should leave the Planck scale invariant.

The Amelino-Camelia (AC) \cite{ame1}, and the Maguejo-Smolin (MS) \cite{mag1} DSR theories are canonical and prototypical, but they are not general DSR theories at  first order Planck length. They are specific examples of DSR proposals. 

Motivated by the DSR proposal, we starting from the boost formula in the $\kappa$-Poincar\'e and DSR theories \cite{Maj,AmBen}, take its leading order in Planck length, and find a generalized boost by adding some free parameters $\beta_i$ to this first order boost ($\beta_i$  are real numbers and $i=0,...,4$.). The generalized boost parametrizes the DSR theories to first order. By using this boost, we obtain differential equations which govern the evolution of the energy and momentums in momentum space. These differential equations are some extensions of the differential equations like $  \frac{d^2p_0}{d\xi^2} - p_0 =0 $, and $ \frac{dp_2}{d\xi}=0 $ in special relativity. With perturbative methods, we obtain solutions of the aforementioned differentials, which lead us to the finite-boost DSR transformations in (3+1) dimensional momentum space.

We show that the Amelino-Camelia (AC) \cite{ame1}, and the Maguejo-Smolin (MS) \cite{mag1}  DSR theories in first order of the Planck length $l_p$, are special cases of our DSR finite-boost transformations. Our study is in the same direction as \cite{cam,Lor,Ive,ame3}, i.e. `beyond special relativity theories' and can be regarded as a logical continuation and extension. It may also prove useful for investigating the geometrical nature of DSR \cite{Jaf}.

Other approaches, such as the generalized uncertainty principle (GUP) is related 
to DSR \cite{Tawf2}.  There is a result of recasting the commutators between $\textbf{p}$  and $ \textbf{x}$ by adding modifications involving the Planck energy \cite{Tawf1, Cap}. As is  well known, the DSR theories are obtained by adding the Planck scale as an invariant to the boost sectors of the Poincar\'e group. Thus, a general expression for the first order DSR finite transformations can assist in finding a consistent GUP theory to first order in the Planck energy.

Observations of quantum gravity effects for astrophysical and cosmological observations are challenging, and finding these tiny effects are notoriously difficult in practice \cite{Fer, abr}. On the theoretical side, there many approaches for including quantum gravity effects (see e.g. \cite{Hos2, ame4} for a review). Finite-boost transformation  DSR theories and classifications of first order DSR theories (by use of a dispersion relation) provide more possibilities for investigating quantum gravity effects in observations.  The freedom comes from the use of adjustable parameters in the finite-boost transformations.

\section{ Lorentz transformations from the differential equations}

The Lorentz transformations in momentum space are
\be   \label{LorentzBoost}    p'_\mu= \Lambda_\mu^{~\nu} p_\nu, \ee
where $ \Lambda_\mu^{~\nu}  $ are the components of the Lorentz matrices. The $p'_\mu$ are the energy and momenta components for a particle in the primed inertial system which moves with velocity $v $ with respect to the unprimed inertial system. For the infinitesimal Lorentz matrices, we have
 \be  \label{InfLorentzmat} \Lambda_\mu^{~\nu} \simeq \delta_\mu^{~\nu} +  \omega_\mu^{~\nu}, \ee 
 where $ \omega^{\mu\nu} $ are the components of the anti-symmetric $ [\omega^{\mu\nu}] $ matrices. In the (1+1)-dimensional case, we have
              \be  \label{omega}  \omega_\mu^{~\nu}= (    \delta_\mu^{~1}  \delta_0^{~\nu}   + 
              \delta_\mu^{~0}  \delta_1^{~\nu}   )\xi.  \ee 
Here, $\xi$ is the rapidity parameter. Thus, the explicit form of the infinitesimal Lorentz transformations are:
          \be \label{InfinitSLT} \left\{\begin{array}{cl}  p'_0 &=  p_0 + p_1\xi , \\
               p'_1 &=  p_1 + p_0\xi, \\
               p'_2 &=  p_2, \\
               p'_3 &=  p_3. \end{array}\right.
               \ee 
From which, we can write the first order differential equations for the $p_0$ and $p_1$ components as       
\be\left\{\begin{array}{cl}   \frac{d p_0}{d \xi}  = p_1  , \\\\
               \frac{d p_1}{d \xi} = p_0.
               \end{array}\right.\ee
The second order differential equations which contains only $p_0$ and $p_1$ are:
\bea  \label{seconorderlorentz}  \left\{\begin{array}{lr}  \frac{d^2p_0}{d\xi^2}- p_0=0,
       \\\\\frac{d^2p_1}{d\xi^2} -p_1=0. \end{array} \right. \eea
The first equation of the differential equations in Eq.~(\ref{seconorderlorentz}) in the primed system has the solution 
 \be \label{p1LorntzT} p_0'(\xi) = C_1 \cosh\xi + C_2\sinh \xi, \ee
 where $C_1$ and $C_2$ are constants, and can be determined from the initial conditions $p'_0(\xi=0)=p_0$ and $\frac{dp'_0}{d\xi}(\xi=0)=p_1 $.
 
 For the $p_2$ and $p_3$ components, the infinitesimal transformations Eq.~(\ref{InfinitSLT}) give the first order differential equations
 \be  \label{p2p3DiffEQLT} \left\{\begin{array}{cl}   \frac{d p_2}{d \xi}=0, \\\\
               \frac{d p_3}{d \xi} = 0.
               \end{array}\right.\ee
 In the primed system, the solution of the first differential equations in Eq.~(\ref{p2p3DiffEQLT}) will be 
 \be \label{p2LorentzT} p_2'(\xi) =p_2+ C_1,\ee
 where $C_1$ is a constant. By using the initial condition $p'_2(\xi=0)=p_2$, one finds that this constant should be zero.
 
 Therefore, from Eq.~(\ref{p1LorntzT}) and Eq.~(\ref{p2LorentzT}), we find the well-known Lorentz transformations in the energy momentum space as
  \bea    \left\{\begin{array}{lr}  p'_0= p_0\cosh\xi + p_1 \sinh\xi,\\
   p'_1= p_1\cosh\xi + p_0 \sinh\xi, \\
               p'_2 =  p_2, \\
               p'_3 =  p_3
   \end{array} \right. \eea
In the next section, we will extend this method for finding solutions of the finite-boost DSR in first order of Planck length transformations which are nonlinear extensions of the system of 
 differential equations in Eq.~(\ref{seconorderlorentz}).
 
 \section{ DSR Theories in first order of Planck length} 
 
 \subsection{DSR Boosts }

The generator of the DSR theories in (3+1) dimensions is given by
              \be \label{ModGen}  N_1=  p_1 \frac{\partial  }{ \partial p_0 } + \Big(\frac{l_p}{2} \textbf{p}^2 +
              \frac{1 - e^{-2l_p p_0} }{2l_p} \Big)\frac{\partial  }{ \partial p_1 } - l_p p_1  \Big( p_j\frac{\partial }{ \partial p_j } \Big).   \ee
    It originates from the $\kappa$-Poincar\'e group  \cite{Maj,AmBen,kow}.         
In leading order of $l_p$, this generator will be 
           \be \label{ModGenFirstOrd}  N'_1=  p_1 \frac{\partial  }{ \partial p_0 } + \Big( p_0 - l_p p_0^2 +\frac{l_p}{2} \textbf{p}^2 \Big)\frac{\partial  }{ \partial p_1 }-
           l_p  p_1 \Big( p_j \frac{\partial  }{ \partial p_j } \Big).   \ee
In fact, this generator represents only one class of  DSRs. In order to include a wider range of theories, we modify this generator by generalizing to preserve parity and time-reversal symmetry: 
 \be \label{3DimGenerator} \Tilde{N}_1=   ( p_1 +  l_p\beta_0 p_0 p_1 ) \frac{\partial  }{ \partial p_0 } + 
 \Big( p_0 +  l_p\beta_1 p_0^2+ l_p\beta_2 \textbf{p}^2 \Big)\frac{\partial  }{ \partial p_1 } + 
 l_p\beta_3 p_1 \Big( p_j \frac{\partial  }{ \partial p_j } \Big)  + l_p\beta_4 \epsilon_{1jk} p_j \frac{\partial  }{ \partial p_k },\ee
 where $\beta_0$, ..., $\beta_4$ are arbitrary real numbers. We parameterize the extended $\kappa$-Poincar\'e symmetry in the first order of the 
 Planck length by these real numbers. One cannot obtain MS-DSR and other DSR theories from Eq.~(\ref{ModGenFirstOrd}), but one can obtain them from Eq.~(\ref{3DimGenerator}). This generalization gives us DSR theories to the first order in the Planck length.  We also have extended freedom in finding more general dispersion relations to match observational results. In other words, Eq.~(\ref{3DimGenerator}) generalizes first order $\kappa$-Poincar\'e symmetry. This extension obeys the group structure and does not violate the relativity principle \cite{cam}.
 
 By using the general generator Eq.~(\ref{3DimGenerator}), we can write the infinitesimal transformation in the first order of the Planck length $l_p$ \cite{cam, Lor2}, as  
\be \label{InfinitDSRTr}  \left\{\begin{array}{cl}  p'_0 &=   p_0 + p_1\xi +  l_p \beta_0  p_0 p_1 \xi, \\\\
               p'_1 &=  p_1 + p_0 \xi + l_p\beta_1 p_0^2 \xi + l_p \beta_2 \textbf{p}^2 \xi + l_p \beta_3 p_1^2 \xi,\\\\ 
    p'_2 &=  p_2 + l_p\beta_3 p_1p_2\xi  - l_p \beta_4 p_0p_3 \xi , \\\\
    p'_3 &=  p_3  +l_p\beta_3 p_1p_3 \xi + l_p \beta_4 p_0p_2 \xi . \end{array}\right.\ee

 \subsection{ Generalization of the first order $\kappa$-Poincar\'e symmetry } \label{Ext-Pion_Algb}
 
 Majid and Ruegg \cite{Maj}, demonstrated that the $ \kappa$-Poincar\'e algebra is a semi-direct product of 
 the classical Lorentz group $SO(1,3)$ acting in a deformed way on the momentum sector. There is a back-reaction of the momentum sector on the Lorentz rotations, while the rotation part of the Lorentz sector is not deformed. Amelino-Camelia and Maguijio and Smolin used similar procedures to obtain the AC-DSR and MS-DSR theories \cite{ame1, ame2, mag1}. They modified only the boosts of the Poincar\'e group and kept the rotations unmodified. These modified boosts act on the momentum space in a nonlinear way.

We continue this procedure for modifying the boost generator by adding the parameters $\beta_i$, but we keep the other generators 
 $ \{ p_0, p_i, M_i \} $ unmodified. We have added only some parameters to Eq.~(\ref{ModGenFirstOrd}) to obtain more general modifications in the resulting new boost in Eq.~(\ref{3DimGenerator}). The new boosts (arbitrary directions) and rotations should be satisfied in an extended group like the $\kappa$-Poincar\'e group to first order. We will check the group properties for the modified boost Eq.~(\ref{3DimGenerator}) in subsection \ref{Comp-boosts}. 
 
Some comments are warranted with respect to the motivation of the new modified boost in Eq.~(\ref{3DimGenerator}) from the boost in Eq.~(\ref{ModGenFirstOrd}). Adding free parameters enlarges the symmetry of the $\kappa$-Poincar\'e group. There is mathematical motivation to study the extension of the $\kappa$-Poincar\'e group and its interesting nonlinear actions.
Moreover, the relevant differential equations involve nonlinear solutions of second order which are important for more general dynamical systems and their symmetries.

To obtain the corresponding algebra for this extended group, we rewrite the general generator in Eq.~(\ref{3DimGenerator}) for any arbitrary direction 
 \be \label{3DimGen_arbitrary} \Tilde{N}_i=   ( p_i +  l_p\beta_0 p_0 p_i ) \frac{\partial  }{ \partial p_0 } + 
 \Big( p_0 +  l_p\beta_1 p_0^2+ l_p\beta_2 \textbf{p}^2 \Big)\frac{\partial  }{ \partial p_i } + 
 l_p\beta_3 p_i \Big( p_j \frac{\partial  }{ \partial p_j } \Big)  + l_p\beta_4 \epsilon_{ijk} p_j \frac{\partial  }{ \partial p_k }.\ee
  We find the commutators of this generator with $p_\mu$ and the commutators of this generator with $M_i$ (the generators of the rotations). We also find the commutators of this generator with itself. This modified 
  algebra is given in the following,   
  \be   [\Tilde{N}_i, p_j ]=     (   p_0   + l_p \beta_1 p_0^2  + l_p \beta_2 \textbf{p}^2  ) \delta_{ij}
      + l_p \beta_3 p_i p_j - l_p \beta_4 \epsilon_{ijk} p_k, \ee
      
      \be   [\Tilde{N}_i, p_0 ]=  p_i   + l_p \beta_0 p_0 p_i.  \ee 
The other commutators are
\be [M_i, \Tilde{N}_j]= \epsilon_{ijk} \Tilde{N}_k - l_p \beta_4 p_0 \epsilon_{ijk} M_k, \ee
      \be \label{BoostCombination} [\Tilde{N}_i, \Tilde{N}_j]= -\epsilon_{ijk} M_k 
      - l_p ( \beta_0 + 2 \beta_1 + 3 \beta_2 - \beta_3  ) \epsilon_{ijk} M_k  - 2 l_p \beta_4 p_0 \epsilon_{ijk} \Tilde{N}_K , \ee
and the rotations of momentums which remain unmodified are
   \be  [M_i, M_j]= \epsilon_{ijk} M_k, \ee    
  \be \hspace{1cm} [M_i, p_j]= \epsilon_{ijk} p_k ,      \hspace{1cm} [M_i, p_0]= 0.  \ee
The algebraic structure of our extended $\kappa$-Poincar\'e group is important for studying symmetries of this group.

  \subsection{ Differential Equations and Finite-Boost Transformations}
 
 From Eq.~(\ref{InfinitDSRTr}), we find differential equations which govern the evolution of energy and momentums in momentum space, and we solve 
 the differential equations for the $p_0$ and $p_1$ components together and for the $p_2$ and $p_3$ components together. For the $p_0$ and $p_1$ components
  we solve the second order equations which are more simple, but for the $p_2$ and $p_3$ components it is best to find the solutions of the first order 
  differential equations.   
  For the $p_0$ and $p_1$ components the first order differential equations are 
    \be  \label{3D_DifEq} \left\{\begin{array}{cl}   \frac{d p_0}{d \xi} &=  p_1 + 
  l_p\beta_0  p_0 p_1 , \\\\
    \frac{d p_1}{d \xi}  &=  p_0 +  l_p\beta_1 p_0^2 + l_p\beta_2 \textbf{p}^2 + l_p\beta_3 p_1^2.  \end{array}\right.\ee 
By introducing the constants
\be    a_0= \beta_0 + \beta_2 , ~~~ a_1= \beta_0 + \beta_2 + \beta_3, ~~~
\textrm{and}~~~ a_3= 2\beta_0 +2\beta_1+ 2\beta_2+ \beta_3,\ee 
we obtain the second order differential equations,
 \bea \label{SecOrderDEBoost}  \left\{\begin{array}{lr}  \frac{d^2p_0}{d\xi^2}= p_0  + l_p a_0 p_0^2  + l_p a_1 (\frac{dp_0}{d\xi})^2 +
     \beta_2 (p_2^2 +p_3^2),
  \\\\
  \frac{d^2p_1}{d\xi^2} =   p_1 + l_p a_3 p_1\frac{dp_1}{d\xi}. \end{array} \right. \eea
For the $ p_2$ and $p_3$ components the differential equations are
\be  \label{p2p3_DifEq} \left\{\begin{array}{cl}  
    \frac{d p_2}{d \xi}  &=  l_p\beta_3 p_1p_2 - l_p\beta_4 p_0p_3, \\\\
    \frac{d p_3}{d \xi}  &=  l_p\beta_3 p_1p_3 + l_p\beta_4 p_0p_2.  \end{array}\right.\ee 
Solutions of the differential equations in Eq.~(\ref{SecOrderDEBoost}) and Eq.~(\ref{p2p3_DifEq}) give us the finite-boost DSR transformations
for all orders of the rapidity parameter $\xi$, but only in the first order of the Planck length. Solutions of these differential equations are given in Appendix \ref{A1}. 

 The finite-boost transformations to the first order in $l_p$ are 
\bea   \label{3DimTransform} \left\{\begin{array}{lr}  p'_0= p_0\cosh\xi + p_1 \sinh\xi 
    \\ \hspace{ 1cm}+ l_p( \alpha_1 p_0^2 + \alpha_2 p_1^2) \sinh^2 \xi + l_p( \alpha_2 p_0^2 + \alpha_1 p_1^2) \cosh^2 \xi 
    \\ \hspace{ 1cm}+ 2 l_p(\alpha_1+ \alpha_2) p_0 p_1 \sinh\xi \cosh \xi - l_p(\alpha_2 p_0^2 + \alpha_1 p_1^2)\cosh \xi
    -  l_p\alpha_3 p_0 p_1 \sinh\xi  \\~~~~~~~ + l_p\beta_2 (p_2^2 + p_3^2 )(\cosh\xi -1),\\\\
   p'_1= p_1\cosh\xi + p_0 \sinh\xi 
   \\ \hspace{1cm}  + l_p \alpha_3  p_0 p_1 \sinh^2 \xi  + l_p \alpha_3  p_0 p_1\cosh^2 \xi
   \\ \hspace{1cm}  + l_p \alpha_3  (p_0^2 +p_1^2)\sinh\xi \cosh \xi   -l_p \alpha_3  p_0 p_1 \cosh \xi - l_p(\alpha_2 p_0^2 +
   \alpha_1 p_1^2)\sinh\xi  \\ ~~~~~~ + l_p\beta_2 (p_2^2 + p_3^2 )\sinh\xi   ,\\\\ 
   p'_2= p_2 + l_p (\beta_3 p_1 p_2 -\beta_4 p_0 p_3 )\sinh\xi + \l_p (\beta_3 p_0 p_2 -\beta_4 p_1 p_3 )(\cosh\xi-1),\\\\
     p'_3= p_3 + l_p (\beta_3 p_1 p_3 +\beta_4 p_0 p_2 )\sinh\xi + \l_p (\beta_3 p_0 p_3 +\beta_4 p_1 p_2 )(\cosh\xi-1), \end{array} \right. \eea
where
         \be  \label{AlphaParameters}
    \alpha_ 1\equiv \frac{\beta_0 + 2\beta_1 -\beta_2 -\beta_3 }{3}, ~~~ 
   \alpha_ 2\equiv \frac{\beta_0 -\beta_1 +2\beta_2 + 2\beta_3 }{3},\; \textrm{and}~~~ \alpha_ 3\equiv \frac{\beta_0 +2\beta_1+2\beta_2+ 2\beta_3 }{3},
   \ee
   for convenience.  These transformations are finite-boost DSR transformations to first order in $l_p$ but for all orders of rapidity $\xi$. We confirm they are the same as the transformations in \cite{Wan}, which have been obtained using commutators.

\subsection{Dispersion Relation and its classifications}

Taking $p_\mu= (p_0, \textbf{p})=(m,\textbf{0})$ for the initial $p_\mu$ in the transformations in  Eq.~(\ref{3DimTransform}), we obtain general expressions for the $\cosh\xi$ and $\sinh\xi$  as 
\be \cosh\xi= \frac{p_0- l_p \alpha_1 \textbf{p}^2 - l_p \alpha_2 p_0}{m},~~~
           \sinh\xi= \frac{ |\textbf{p}|- l_p  \alpha_3  p_0 |\textbf{p}|}{m}.\ee
Here, $m$ is the rest mass of the particle. Using $ \cosh^2\xi- \sinh ^2\xi=1 $, we find the corresponding dispersion relation for the transformations
of Eq.~(\ref{3DimTransform})  to first order in $l_p$ as
  \be  \label{DispRelAlpha} p_0^2- \textbf{p}^2 - 2 l_p \alpha_2 p_0^3 + 2 l_p ( \alpha_3 - \alpha_1 )p_0 \textbf{p}^2 =m^2. \ee
Expressing this dispersion relation in terms of $\beta_i $ yields 
  \be   p_0^2- \textbf{p}^2 - 2 l_p \Big(\frac{\beta_0 - \beta_1 + 2\beta_2 +2 \beta_3 }{3} \Big)p_0^3 + 2 l_p (\beta_2+ \beta_3) p_0 \textbf{p}^2 =m^2. \ee
 Thus, if we know the infinitesimal transformations for every given DSR in first order of Planck length as in Eq.~(\ref{InfinitDSRTr}) we can read 
 the $ \beta_0 $,...,$\beta_4$ parameters from them. Then we can compute $ \alpha_1$ to $\alpha_3 $ parameters in Eq.~(\ref{AlphaParameters}). By putting 
 them in Eq.~(\ref{3DimTransform}), we can construct the corresponding finite-boost transformations for the given infinitesimal transformations. Moreover, 
 the corresponding modified dispersion relation can be found from Eq.~(\ref{DispRelAlpha}). This approach highlights the fact that an infinite number of 
 the finite-boost DSR transformations in first order of Planck length theories are possible \cite{kow}. 
 
 We can classify the finite-boost DSR transformations in first order with respect to the different types of modified dispersion relations. In first order of $l_p$, the modified dispersion relation is given by Eq.~(\ref{DispRelAlpha}) which contains the $ p_0 p_1^2$ and the $ p_0^3 $ additional terms. The dispersion relation for MS-DSR in Eq.~(\ref{MSDiperRel}) contains these two additional terms as well, which we regard as a first type of classification. The dispersion relation for AC-DSR in Eq.~(\ref{ACDispRel}) to the first order in $l_p$ contains only the coupling $ p_0 p_1^2$ additional term, which is the second type. The third type of the classification contains only the $ p_0^3 $ additional term. The fourth type, which is our final type of classification has no additional term but the transformations are different from Lorentz.

\subsection{ Composition of Boosts and Wigner Rotation}\label{Comp-boosts}

For testing the group properties of the boosts in Eq.~(\ref{3DimTransform}), we check that the composition of two boosts is a new different boost similar to Lorentz. The composition of boosts for two parallel boosts should be a new boost, and for two perpendicular boosts we should 
      have a new boost with a rotation (Wigner rotation). These properties have been checked in \cite{Wan} for the generalized boost in Eq.~(\ref{3DimGenerator}). Here, we check and confirm in detail some of these properties.

      First we check combination property for the infinitesimal transformations Eq.~(\ref{InfinitDSRTr}). We assume another infinitesimal transformations like Eq.~(\ref{InfinitDSRTr}), but from $p'_{\mu}$ to $ p''_{\mu}$ with rapidity parameter 
      $\xi'$ in the same direction which is given by 
      
       \be \label{InfinitTRprim}  \left\{\begin{array}{cl}  p''_0 &=   p'_0 + p'_1 \xi' +  l_p \beta_0  p'_0 p'_1 \xi', \\\\
               p''_1 &=  p'_1 + p'_0 \xi' + l_p \beta_1 p_0^{\prime 2} \xi' + l_p \beta_2 {\textbf{p}}^{\prime 2} \xi' + l_p \beta_3 p_1^{\prime 2} \xi',\\\\     p''_2 &=  p'_2 + l_p\beta_3 p'_1 p'_2\xi'  - l_p \beta_4 p'_0 p'_3 \xi' , \\\\
             p''_3 &=  p'_3  +l_p\beta_3 p'_1 p'_3 \xi' + l_p \beta_4 p'_0 p'_2 \xi'.\end{array}\right.\ee
    
    Combinations of the infinitesimal transformations in Eq.~(\ref{InfinitDSRTr}) and Eq.~(\ref{InfinitTRprim}) will give us other similar infinitesimal transformations  from $p_{\mu}$ to $ p''_{\mu}$ in the same direction with rapidity parameter $ \xi''$, which is the sum of two rapidity parameters, 
    \be  \label{rapiditySUM}  \xi''=\xi + \xi'.\ee
In the finite-boosts transformations case, as given by Eq.~(\ref{3DimTransform}), we have expressions that include $\cosh{\xi}$, $\sinh{\xi}$, 
$\cosh^2{\xi}$, $\sinh^2{\xi}$, and $\cosh{\xi} \sinh{\xi}$, which yield the same transformations after two successive boosts in the same direction and the rapidity parameters satisfy  Eq.~(\ref{rapiditySUM}) as in the infinitesimal transformations case. 

For two perpendicular boosts, as mentioned, we have a new boost and a rotation. In special relativity this rotation is the well-known Wigner rotation $ \theta_W $ as discussed in \cite{Wig, Ferraro} and is given by 
\be    \tanh{\theta_W} =-\frac{\gamma \gamma' \xi \xi' }{ \gamma + \gamma' }.\ee
In DSR theories, the rotation part of the Poincar\'e group are unmodified and only the boosts are changed.  Therefore 
no correction will be added to the Wigner rotation \cite{Wan}.

The interesting consequences of the Wigner rotation in special relativity have been studied before in \cite{Wig,Ferraro, Ma}. An important effect is related to the spin of a composite system such as a proton which is a composite system of quarks. The sum of the spins for a composite system can violate Lorentz invariance. In fact, spin is related to the Poincar\'e group, or, to the $\kappa$-Poincar\'e group and its extensions, such as the extension in subsection \ref{Ext-Pion_Algb}. The spin $S$ of a moving particle with mass $m $ can be defined
by transforming its Pauli-Lubinski 4-vector $ w_{\mu}=(1/2) \epsilon_{\nu \rho\sigma \mu } J^{\rho \sigma} p^{\nu} $ to its rest frame by a rotationless DSR boost $D(p) $, where $ D(p)p = (m, \textbf{0}) $, and  $ (0, \textbf{S}) = D(p) w/m $. Under an arbitrary DSR transformation
$ \Tilde{\Lambda}  $, the spin and momentum of a particle will be transformed via  
\be      \textbf{S}'= R_w(\Tilde{\Lambda}, p )\textbf{S}  ,  ~~ p'= \Tilde{\Lambda} p,\ee
where    \be \label{WigDSR}  R_w(\Tilde{\Lambda}, p )=  D(p') \Tilde{\Lambda}  D^{-1} (p),  \ee
is the Wigner rotation \cite{Ma}. By using Eq.~(\ref{WigDSR}), we can show that the Wigner rotation for DSR is equal to the Lorentz case.

In the primed moving frame, the proton is boosted with a rotationless DSR transformation along its spin direction.  Each
quark spin will be affected by a Wigner rotation, and these rotations can change the vector spin sum of the 
quarks and antiquarks. This changing of the spin sum has  consequences for the parton model of the proton. 
However, the proton spin in the moving frame will be the same as in the
rest frame \cite{Ma}. Since the Wigner rotation in DSR is the same as the special relativistic case, there are no new consequences (i.e. no new kinds of spin) of the Wigner rotations besides the usual special relativistic effects.

The group of the DSR theories is the $\kappa$-Poincar\'e group \cite{Maj,AmBen,kow}, and our extension of this group in 
      Eq.~(\ref{3DimGenerator}) in first order Planck length, is an extension of the $\kappa$-Poincar\'e symmetry. 
    The commutators between generators for every two boosts in one direction, or in two different directions satisfy Eq.~(\ref{BoostCombination}), 
    which shows that combinations of two boosts will give another boost, or a boost and a rotation. We do not check these properties in full 
      detail; however, they can be tested with the transformations in Eq.~(\ref{3DimTransform}) by long but straightforward calculations. 
      
      We have tested the group property and find the generator in Eq.~(\ref{3DimGenerator}) satisfies an extension of the $\kappa$-Poincar\'e algebra as
      given in subsection \ref{Ext-Pion_Algb}.  This is assurance that the finite-boost transformations in Eq.~(\ref{3DimTransform}) are not just 
      reparameterizations of the compact Lie groups as discussed in \cite{hos}.

 \section{ Special Examples of the Finite-Boost DSR Transformations}
 
 In this section, we provide some examples of the finite-boost DSR transformations in first order of Planck length by finding finite-boost transformations and their corresponding
 dispersion relations to first order in $l_p$ from the general formalism. The first two examples 
 are the well-known MS and AC-DSR theories, which help confirm the validity of the formalism. The last final two examples are of particular importance, because they have special form dispersion relations.  
 
 \subsection{MS-DSR} 

 For the MS-DSR we have 
  \be \label{MS-const1}  \beta_0=-1,~ \beta_1=0,~\beta_2=0,~\beta_3=-1,\,\textrm{and}~ \beta_4=0,  \ee
 and we can compute the constants $\alpha_1$ to $\alpha_3$ from the relations in Eq.~(\ref{AlphaParameters}) as
  \be  \label{MS-const2} \alpha_1=0, ~~ \alpha_2=-1,\, \textrm{and}~ \alpha_3= -1.\ee 
  By putting these values in Eq.~(\ref{DispRelAlpha}) we find the dispersion relation of the MS-DSR to the first order in $l_p$,  
   \be \label{MSDiperRel}  p_0^2- \textbf{p}^2 + 2 l_p p_0^3 -2 l_p p_0 \textbf{p}^2  =m^2. \ee
   Transformations for this  DSR in first order are given in Appendix \ref{A2}.

 \subsection{AC-DSR} 
      
      For the AC-DSR  we have 
       \be \label{AC-parametrs1}   \beta_0=0,~ \beta_1=-1,~\beta_2=\frac{1}{2},~\beta_3=-1,\,\textrm{and}~ \beta_4=0,\ee 
       and we can compute the constants $\alpha_1$ to $\alpha_3$  by use of the relations in Eq.~(\ref{AlphaParameters}) as
       \be  \label{AC-parametrs2} \alpha_1=-\frac{1}{2}, ~~ \alpha_2=0, \,\textrm{and}~ \alpha_3= -1.\ee 
By using these values, we find dispersion relation of the AC-DSR in the first order of $l_p$ as
  \be \label{ACDispRel} p_0^2- \textbf{p}^2 - l_p p_0 \textbf{p}^2  =m^2, \ee
  and transformations for this DSR are given in Appendix \ref{A3}.

 \subsection{Dispersion Relation with the Cubic Term }
 
Using the conditions $\alpha_1 = \alpha_3  $, and $\alpha_2\neq 0$ in Eq.~(\ref{DispRelAlpha}) one finds a modified dispersion relation with a cubic term, 
  \be \label{cubic-Dis_Rel }  \ p_0^2- \textbf{p}^2 - 2 l_p \alpha_2 p_0^3 =m^2. \ee
 Transformations for this type of DSR are in Appendix \ref{A4}.

\subsection{Special Relativistic Dispersion Relation }

The final interesting example utilizes $\alpha_1 = \alpha_3   $,  and $\alpha_2= 0$ in Eq.~(\ref{DispRelAlpha}). These conditions lead us to the unmodified special relativistic dispersion relation
\be  \label{SRDisRel} p_0^2- \textbf{p}^2 =m^2. \ee
 In terms of the $\beta_i$ parameters these conditions will be
     \be  \label{SRCond} \beta_2=-\beta_3  ~\textrm{and},~  \beta_0= \beta_1, \ee
    which is valid for many general DSR theories in first order of Planck length.  
     By putting the conditions of Eq.~(\ref{SRCond}) into Eq.~(\ref{AlphaParameters}) we find  
    \be  \alpha_1=B, ~~ \alpha_2=0, ~~ \alpha_3 = B.\ee 
    Here, we have taken $B \equiv\beta_0$, which is the free parameter. There also two other free parameters $\beta_2$ and $\beta_4$, which are taken to be zero for the following transformations. 
    By using the values of $ \alpha_1$ to $\alpha_3 $ parameters in Eq.~(\ref{3DimTransform}), we find the following transformations
        \bea \label{SRDisRelTransf}
     \left\{\begin{array}{lr}  p'_0= p_0\cosh\xi + p_1 \sinh\xi 
    \\ \hspace{2.4cm}
    + l_p B \Big[  ( p_0  \sinh\xi + p_1 \cosh \xi)^2 -p_1^2\cosh \xi -   p_0 p_1 \sinh\xi  \Big],    \\
    p'_1= p_1\cosh\xi + p_0 \sinh\xi 
   \\ \hspace{2.4cm} + l_p B \Big[  p_0 p_1 (\sinh^2\xi + \cosh^2\xi)\\ \hspace{3.4cm} 
   +(p_0^2 +p_1^2)\sinh\xi \cosh \xi - p_0 p_1 \cosh \xi - p_1^2 \sinh\xi\Big], \\
     p'_2= p_2,\\
     p'_3= p_3.\end{array} \right.
   \eea
  These transformations are the finite-boost DSR transformations in the first order of Planck length $l_p$, with an unmodified special relativistic dispersion relation. Despite this, they are not the usual Lorentz transformations. It is surprising that the special relativistic dispersion relation accompanies these non-Lorentz transformations. 
   
   \subsection{Dispersion Relation Family }\label{SRfamily}
   The transformations Eq.~(\ref{SRDisRelTransf}) which corresponds to the dispersion relation for the special relativity are not unique. In condition Eq.~(\ref{SRCond}), there are three free parameters: $\beta_0$, $\beta_2$, and $\beta_4 $. A second example of the transformations which have the special relativistic dispersion relation in the first order of the Planck length is found by taking $\beta_4=1$ instead while keeping $\beta_0=B$ and $\beta_2=0$ as before. These transformations are given in Appendix \ref{A5}.
   
   We refer to these transformations as the second finite-boost DSR transformations with special relativistic dispersion relation. Of course, for other choices of the free parameters one can find other transformations. Thus, one has a family of transformations of the DSR theories in first order. For the dispersion relation with a cubic term, one has another similar family.

    \section{Observational Consequences of the first order DSR}\label{V}

 The proposal for observation of quantum gravity effects in gamma ray bursts was given initially by Amelino-Camelia and his colleagues in \cite{ame5}. They assumed a deformed dispersion relation for the photons in the form of
  \be    p_1^2=E^2 \Big[   1 + f \Big( \frac{E}{E_{QG}} \Big) \Big], \ee
  where $ E_{QG} $ is an effective quantum gravity scale, and $f$ is a model dependent function of $ E/E_{QG} $. One can find an upper limit on the scale with observations and it may be different from the Planck energy $ M \simeq 1.22 \times 10^{19}  \textrm{GeV}$.  
  This dispersion relation to first order in $ E/E_{QG} $  is
  \be \label{DisRelNat}  p_1^2=E^2 \Big(   1 + \eta \frac{E}{E_{QG}}  \Big),   \ee
 where $\eta $   is some number. The proposal is that the velocity of the gamma ray bursts should be different from $c$, which is given by 
 \be \label{LinearVgamma}     v=  \frac{\partial E}{ \partial p_1} = c \Big ( 1- \eta \frac{E}{E_{QG}}   \Big). \ee
 Moreover, a signal which is coming from distance $D$ with energy $E$ will be received with time delay 
   \be    \Delta t_{LIV} = \eta \  \frac{E}{E_{QG}}\frac{D}{c}, \ee
different than the ordinary case with the speed $c$. A linear correction to the velocity of high energy photons like Eq.~(\ref{LinearVgamma}) has been suggested in many papers for testing quantum gravity effects \cite{Fer,abr, hu, lan, joh, hos, joh2,Pan}. 

We find the velocity of photons and time delay for the photons in our formalism for finite-boost DSR transformations. 
By using the dispersion relation in Eq.~(\ref{DispRelAlpha}), the velocity of the photons will be 
\be \label{VlinearExperim}  v\simeq  c \Big[1 + (\alpha_1 + \alpha_2 - \alpha_3  ) l_p E \Big], \ee 
and the time delay will be
 \be  \label{TimeDelayMSR} \Delta t_{LIV}  =  (\alpha_1 + \alpha_2 - \alpha_3 ) \frac{ l_p E D}{c}. \ee
 As can be seen, the $\eta$ parameter in Eq.~(\ref{DisRelNat}) has been replaced by $ (\alpha_1 + \alpha_2 - \alpha_3 )$ in the finite-boost DSR transformations. To be clear, the distance $D$ is a cosmological distance that depends on the redshift $ z$,  matter density  $\Omega_m $, cosmological constant 
 density $\Omega_{\Lambda}$, and Hubble constant $ H_0$. The time delay is
\be  \label{TimeDelayCosm}  \Delta t_{LIV} = \frac{ (\alpha_1 + \alpha_2 - \alpha_3 ) l_p  E  }{H_0 } \int_0^z  \frac{(1+ z') dz'}{ \sqrt{1+ \Omega_m (1+ z')^3 + \Omega_{\Lambda} } }.   \ee
Here, we take  $\Omega_m=0.3 $, $\Omega_{\Lambda}=0.7 $, $ H_0= 72~ \textrm{km} ~\textrm{s}^{-1} ~ \textrm{Mpc}^{-1}  $
        for the cosmological parameters   \cite{joh2}.
 
    The time delay in Eq.~(\ref{TimeDelayMSR}) and Eq.~(\ref{TimeDelayCosm}) is the only Lorentz invariance
    violating (LIV) term for time delay between a high energy astrophysical photon source and receiver. The general expression for time delay is 
     \be   \Delta t= \Delta t_{LIV} +   \Delta t_{int}  +  \Delta t_{spe} + \Delta t_{DM}  + \Delta t_{grav},   \ee
where $\Delta t_{int}$ is the time delay due to the fact that photons with high and low energies do not leave the source simultaneously. $\Delta t_{spe}$ is a result of the special relativistic effects if the photons have non-zero mass and $\Delta t_{DM}$ is due to dispersion by the line-of-sight of free electron content, which is non-negligible especially for low energy photons. Finally, $\Delta t_{grav}$ is due to the gravitational potential contribution along the photons propagation paths for possible violation of Einstein's equivalence principle \cite{Pan}.

   \subsection{ Special Relativistic dispersion  relation and null results in observations for QG Effects}
   
    For $\alpha_1 = \alpha_3  $, and $\alpha_2= 0$, which gives the special relativistic dispersion in Eq. (\ref{SRDisRel}), the velocity of the photons will be the unchanged special relativistic $c$, and the time delay in Eq.~(\ref{TimeDelayMSR}) will be zero. Therefore, the transformations in Eq.~(\ref{SRDisRelTransf}) (and also all transformations in the special relativistic dispersion relation family such as Eq.~(\ref{SRDisRelTransfN2})) are different from the Lorentz transformations in first order of the Planck length, but the velocity of the photons are the same as special relativity. These photons will be received in the same time as in the usual special relativistic case. Simply put, as has been done e.g. in \cite{Fer, abr}, measuring the velocity and time delay of GRB photons is not sufficient for investigating quantum gravity effects to first order.

  \section{ Conclusion}

  Finding general expressions for DSR to first order is an interesting and important matter both mathematically and physically. In this paper we started with the boost from the $\kappa$-Poincar\'e group, adding free parameters $\beta_i$, useful for generalization. The new generalized boost gives finite-boost transformations which allow us to obtain the general expressions for DSR theories in the first order of the Planck length. After finding the dispersion relation for these general finite-boost transformations, we classify the resulting DSR theories to four types. The well-known MS and AC theories are the first two types of this classification. The second two types are the DSR theories with a cubic term in the dispersion relation and the DSR theories with special relativistic dispersion. 
  
  There are important observational consequences of these different DSR types.  Perhaps the most notable consequence is the expectation that the speed of light should be different from the usual speed $c$. In fact,
  the speed of high energy photons and low energy photons should be different and depends on their energy content. In this study, we have found an  interesting possibility where DSR theories with transformations different from Lorentz transformations, can also correspond to an unmodified special relativistic dispersion relation.  Thus, any search for quantum gravity effects in astrophysical and cosmological observations which give a special relativistic result can also be interpreted as consistent with DSR.  A null result is not enough for investigating corresponding quantum gravity effects to first order in Planck length. 
  
   There are two possible extensions and applications of these finite-boost DSR transformations.   First, the solutions of the extended second order differential equations can be investigated for a better understanding of their dynamics and behavior \cite{Mig, Ghos}. Second, the possible extensions of the Poincar\'e or $\kappa$-Poincar\'e group may be fruitful for investigations (e.g. dual DSR theory \cite{Magp1, Magp2})  of quantum gravitational effects.

 \acknowledgments 

Funding by the FY2018-SGP-1-STMM Competitive Research Grant No. 090118FD5350 at Nazarbayev University is appreciated, as well as the state-targeted program ``Center of Excellence for Fundamental and Applied Physics" (BR05236454) by the Ministry of Education and Science of the Republic of Kazakhstan.

 \appendix

   \section{Appendixes} 
   
   \subsection{ Finding  solutions of the differential equations } \label{A1}
  
 For obtaining the finite-boost transformations, first we solve the differential equations for the $p_0$ and $p_1$ components. In the primed inertial system, the form of Eqs.~(\ref{SecOrderDEBoost}) are the same. We introduce the perturbative solutions of these differential equations in first order of the Planck length $l_p$ as
\bea \label{Genralp0p1}   \left\{\begin{array}{lr}  p'_0= p_0\cosh\xi + p_1 \sinh\xi + l_p A(\xi)  ,
\\\\
   p'_1= p_1\cosh\xi + p_0 \sinh\xi + l_p D(\xi) . \end{array} \right. \eea
   The unknown function $A(\xi)$ should also satisfy the following additional differential equation
   \be      \frac{d^2A}{d\xi^2} -A = (a_0 p_1^2 + a_1 p_0^2 ) \sinh^2 \xi+
   (a_0 p_0^2 + a_1 p_1^2 )\cosh^2 \xi + 2(a_0 + a_1) p_0 p_1 \sinh\xi \cosh \xi. \ee
   The general solution for $ A(\xi)$ is 
   \be \label{27}  A(\xi)= \lambda_1 \sinh^2 \xi + \lambda_2\cosh^2 \xi + \lambda_3\sinh\xi \cosh \xi + \lambda_4\cosh \xi +  \lambda_5\sinh\xi - \beta_2 (p_2^2 +p_3^2),   \ee
   where $\lambda_1 = \frac{2a_0-a_1}{3} p_0^2 +  \frac{2a_1-a_0}{3} p_1^2 $, $ \lambda_2 = \frac{2a_1-a_0}{3} p_0^2 +  \frac{2a_0-a_1}{3} p_1^2 $, and $ \lambda_3= \frac{2(a_0+a_1)}{3} p_0 p_1$. 
   To find $ \lambda_4 $, and $ \lambda_5 $ multipliers we use the initial conditions $p'_0(0)=p_0$, and $\frac{dp'_0}{d\xi}(0)=p_1 +  l_p \beta_0 p_0 p_1$, which yield
   $\lambda_4=-\lambda_2  + \beta_2 ( p_2^2 + p_3^2)$, and $ \lambda_5=\beta_0 p_0 p_1- \lambda_3$. The other unknown function $D(\xi)$ can be found in a similar way. By putting 
   $A(\xi)$ and $D(\xi)$ in Eq.~(\ref{Genralp0p1}) we can find the transformations for the $p_0$ and $p_1$ components.
   
    For the $p_2$ and $p_3$ components, we take
      \bea    \left\{\begin{array}{lr}  p'_2= p_2 + l_p A_1\sinh\xi + \l_p A_2 \cosh\xi + A_3  ,\\\\
        p'_3= p_3 + l_p A_4\sinh\xi + \l_p A_5 \cosh\xi + A_6, \end{array} \right. \eea
        where $A_1$, ..., $A_6$, are unknown functions of $ p_0,..., p_3 $. By putting these solutions in the differential equations for $p_2$ and $p_3$ in 
    Eq.~(\ref{p2p3_DifEq}), we can find the unknown functions $A_1$, ..., $A_6$, which lead us to the transformations for the $p_2$ and $p_3$ components.

      \subsection{ MS transformations in the first order of the Planck length}\label{A2}

 By using the values of constants in Eq.~(\ref{MS-const1}) and Eq.~(\ref{MS-const2}) in Eq.~(\ref{3DimTransform}), we find the MS-DSR transformations to first order in $l_p$  as

     \bea
     \label{MSFirstOrder} \left\{\begin{array}{lr}  p'_0= p_0\cosh\xi + p_1 \sinh\xi 
    \\\hspace{1.2cm}
    -l_p p_1^2 \sinh^2 \xi - l_p p_0^2 \cosh^2 \xi  
    \\ \hspace{1.2cm} -2 l_p p_0 p_1 \sinh\xi \cosh \xi + l_p p_0^2 \cosh \xi +  l_p p_0 p_1 \sinh\xi ,
    \\\\
   p'_1= p_1\cosh\xi + p_0 \sinh\xi 
   \\ \hspace{1.2cm}
    -l_p p_0 p_1 \sinh^2 \xi - l_p p_0 p_1 \cosh^2 \xi 
    \\ \hspace{1.2cm}- l_p(p_0^2 + p_1^2) \sinh\xi \cosh \xi + l_p p_0 p_1 \cosh \xi +  l_p p_0^2 \sinh\xi,\\\\
     p'_2= p_2 +\l_p p_0 p_2 \cosh\xi - l_p p_1 p_2\sinh\xi +  l_p p_1 p_2,\\\\
     p'_3= p_3 + \l_p p_0 p_3 \cosh\xi - l_p p_1 p_3\sinh\xi  +l_p p_1 p_3. \end{array}\right.
   \eea
  
 If we expand the expressions for the transformations and dispersion relation of the MS-DSR theory which has been given in \cite{mag1}, we will find the same expressions as in Eq.~(\ref{MSFirstOrder}) and  Eq.~(\ref{MSDiperRel}) to first order in the Planck length.

     \subsection{ AC transformations in the first order of the Planck length}\label{A3}
     
     Using the parameters in Eq.~(\ref{AC-parametrs1}) and  Eq.~(\ref{AC-parametrs2}), we find transformations for the AC-DSR to the first order in $l_p$ as 
     
    \bea
     \left\{\begin{array}{lr}  p'_0= p_0\cosh\xi + p_1 \sinh\xi 
    \\ \hspace{1.2cm}
    -\frac{1}{2} l_p p_0^2 \sinh^2 \xi - \frac{1}{2} l_p p_1^2 \cosh^2 \xi  
    \\ \hspace{1.2cm} -l_p p_0 p_1 \sinh\xi \cosh \xi + \frac{1}{2} l_p p_1^2 \cosh \xi +  l_p p_0 p_1 \sinh\xi , \\~~~~~~~~~~ + \frac{1}{2} l_p (p_2^2 + p_3^2 )(\cosh\xi -1)
    \\\\
   p'_1= p_1\cosh\xi + p_0 \sinh\xi 
   \\ \hspace{1.2cm}
    -l_p p_0 p_1 \sinh^2 \xi - l_p p_0 p_1 \cosh^2 \xi 
    \\ \hspace{1.2cm}- l_p(p_0^2 + p_1^2) \sinh\xi \cosh \xi + l_p p_0 p_1 \cosh \xi +  \frac{1}{2} l_p \textbf{p}^2 \sinh\xi, 
    \\\\
     p'_2= p_2 +\l_p p_0 p_2 \cosh\xi - l_p p_1 p_2\sinh\xi +  l_p p_1 p_2,\\\\
     p'_3= p_3 + \l_p p_0 p_3 \cosh\xi - l_p p_1 p_3\sinh\xi +  l_p p_1 p_3. \end{array} \right.
     \eea
     These transformations and dispersion relation Eq.~(\ref{ACDispRel}) are in agreement with AC-DSR theory \cite{ame1}.
     
    \subsection{ Transformations for the DSR with Cubic Term }\label{A4}

For the DSR with the cubic term, the conditions $\alpha_1 = \alpha_3  $, and $\alpha_2\neq 0$   are equivalent to the conditions 
  \be   \label{CubicTermCond}    \beta_2=-\beta_3,  ~\textrm{and}~  \beta_0\neq \beta_1, \ee 
  for the $\beta_i$ parameters, which is valid for many DSR theories. There are also two other free parameters $\beta_2$ and $\beta_4$, 
  which are taken to be zero for the following transformations. 
  By putting these conditions in Eq.~(\ref{AlphaParameters}) we find
         
     \be 
    \alpha_ 1= \frac{\beta_0 + 2\beta_1 }{3}, ~~~ \alpha_ 2= \frac{\beta_0-\beta_1 }{3}, \,~~~\textrm{and}~~~ \alpha_ 3= \frac{\beta_0 + 2\beta_1 }{3}.
     \ee
    As a specific case, we can take $ \beta_0=0 $,  and  $ B\equiv \beta_1/3$. From which, we have $ \alpha_1 =2B $,  $ \alpha_2=-B $, and $ \alpha_3=2B $. 
    Therefore, an example of the transformations for this type in first order of Planck length is 
    \bea \label{CubTermTransf}
      \left\{\begin{array}{lr}  p'_0= p_0\cosh\xi + p_1 \sinh\xi 
    \\ \hspace{ 1cm}+ l_p B( 2 p_0^2 - p_1^2) \sinh^2 \xi + l_p B( 2 p_1^2 - p_0^2) \cosh^2 \xi 
    \\ \hspace{ 1cm}+ 2 l_p B p_0 p_1 \sinh\xi \cosh \xi - l_p B( 2 p_1^2 - p_0^2)\cosh \xi - 2 l_p B  p_0 p_1 \sinh\xi ,
    \\\\
   p'_1= p_1\cosh\xi + p_0 \sinh\xi 
   \\ \hspace{1cm} + 2 l_p B  p_0 p_1 \sinh^2 \xi + 2l_p B  p_0 p_1\cosh^2 \xi
   \\ \hspace{1cm}  + 2 l_p B (p_0^2 +p_1^2)\sinh\xi \cosh \xi   - 2 l_p B p_0 p_1 \cosh \xi + l_p B( 2p_1^2 -  p_0^2)\sinh\xi,
   \\
     p'_2= p_2,\\
     p'_3= p_3.
   \end{array} \right.
   \eea
    
     \subsection{ Transformations for Special Relativistic Dispersion Relation Family }\label{A5}
     
      The transformations for the special relativistic dispersion relation family which has been discussed in subsection \ref{SRfamily}, are 
      
     \bea \label{SRDisRelTransfN2}
     \left\{\begin{array}{lr}  p'_0= p_0\cosh\xi + p_1 \sinh\xi 
    \\ \hspace{2.4cm}
    + l_p B \Big[  ( p_0  \sinh\xi + p_1 \cosh \xi)^2 -p_1^2\cosh \xi -   p_0 p_1 \sinh\xi  \Big],
    \\\\
   p'_1= p_1\cosh\xi + p_0 \sinh\xi 
   \\ \hspace{2.4cm} + l_p B \Big[  p_0 p_1 (\sinh^2\xi + \cosh^2\xi)\\ \hspace{3.4cm} 
   +(p_0^2 +p_1^2)\sinh\xi \cosh \xi - p_0 p_1 \cosh \xi - p_1^2 \sinh\xi\Big], \\\\
     p'_2= p_2 - l_p p_0 p_3 \sinh\xi -l_p p_1 p_3 (\cosh\xi-1),\\\\
     p'_3= p_3 + l_p  p_0 p_2 \sinh\xi + l_p p_1 p_2 (\cosh\xi-1). 
     \end{array} \right.
     \eea

\end{document}